\documentclass[sigconf]{acmart}

\AtBeginDocument{%
  \providecommand\BibTeX{{%
    \normalfont B\kern-0.5em{\scshape i\kern-0.25em b}\kern-0.8em\TeX}}}

\copyrightyear{2024}
\acmYear{2024}
\setcopyright{acmlicensed}\acmConference[WWW '24 Companion]{Companion
Proceedings of the ACM Web Conference 2024}{May 13--17, 2024}{Singapore,
Singapore}
\acmBooktitle{Companion Proceedings of the ACM Web Conference 2024 (WWW
'24 Companion), May 13--17, 2024, Singapore, Singapore}
\acmDOI{10.1145/3589335.3648329}
\acmISBN{979-8-4007-0172-6/24/05}

\usepackage{float}                  
\usepackage{subfig}                 
\usepackage{overpic}  
\usepackage{color}
\usepackage{multirow}
\usepackage{enumitem}
\begin{document}


\title{PPM : A Pre-trained Plug-in Model for Click-through Rate Prediction}


\author{Yuanbo Gao}
\orcid{0000-0002-0667-2487}
\affiliation{
    \institution{Jingdong Group}
    \city{Beijing}
    \country{China}
}
\email{gaoyuanbo1@jd.com}
\authornotemark[1]

\author{Peng Lin}
\orcid{0009-0007-3844-460X}
\affiliation{
    \institution{Jingdong Group}
    \city{Beijing}
    \country{China}
}
\email{linpeng47@jd.com}
\authornote{Both authors contributed equally to this research.}\authornote{Corresponding author.}

\author{Dongyue Wang}
\orcid{0009-0002-4775-6211}
\affiliation{
    \institution{Jingdong Group}
    \city{Beijing}
    \country{China}
}
\email{wangdongyue@jd.com}

\author{Feng Mei}
\orcid{0009-0002-9442-2548}
\affiliation{
    \institution{Jingdong Group}
    \city{Beijing}
    \country{China}
}
\email{meifeng6@jd.com}

\author{Xiwei Zhao}
\orcid{0000-0002-9382-6041}
\affiliation{
    \institution{Jingdong Group}
    \city{Beijing}
    \country{China}
}
\email{zhaoxiwei@jd.com}

\author{Sulong Xu}
\orcid{0000-0003-0345-334X}
\affiliation{
    \institution{Jingdong Group}
    \city{Beijing}
    \country{China}
}
\email{xusulong@jd.com}

\author{Jinghe Hu}
\orcid{0009-0002-1546-5807}
\affiliation{
    \institution{Jingdong Group}
    \city{Beijing}
    \country{China}
}
\email{hujinghe@jd.com}

\renewcommand{\shortauthors}{Yuanbo Gao et al.}

\begin{abstract}
 Click-through rate (CTR) prediction is a core task in recommender systems. Existing methods (IDRec for short) rely on unique identities to represent distinct users and items that have prevailed for decades. On one hand, IDRec often faces significant performance degradation on cold-start problem; on the other hand, IDRec cannot use longer training data due to constraints imposed by iteration efficiency. Most prior studies alleviate the above problems by introducing pre-trained knowledge(e.g. pre-trained user model or multi-modal embeddings). However, the explosive growth of online latency can be attributed to the huge parameters in the pre-trained model. Therefore, most of them cannot employ the unified model of end-to-end training with IDRec in industrial recommender systems, thus limiting the potential of the pre-trained model.

  To this end, we propose a \textbf{p}re-trained \textbf{p}lug-in CTR \textbf{m}odel, namely PPM. PPM employs multi-modal features as input and utilizes large-scale data for pre-training. Then, PPM is plugged in IDRec model to enhance unified model's performance and iteration efficiency. Upon incorporating IDRec model, certain intermediate results within the network are cached, with only a subset of the parameters participating in training and serving. Hence, our approach can successfully deploy an end-to-end model without causing huge latency increases. Comprehensive offline experiments and online A/B testing at JD E-commerce demonstrate the efficiency and effectiveness of PPM.
  
\end{abstract}





\maketitle

\section{Introduction}
As one of the largest B2C e-commerce platforms in China, JD.com\footnote{\url{https://www.jd.com/}} serves hundreds of millions of active customers. The accurate capturing of user interests assumes heightened significance as they browse the JD App. The traditional methods utilize unique identities to represent distinct users and items have been state-of-the-art and dominated the recommendation systems literature for decades~\cite{sun2019bert4rec}. 

However, when confronted with unfamiliar items that lack interaction features like clicks and orders, the efficacy of ID features in accurately characterizing both users and items falters in such scenarios~\cite{yuan2020parameter}. Another issue is how to strike a balance between training data and iteration efficiency. Despite the abundance of artificial feedback data in industrial domains, it is customary to employ a limited subset of recent data for model training to improve iteration efficiency. Nevertheless, this could potentially diminish the model's performance by limiting the utilization of training data.

The most common way to address these issues is to leverage pre-train knowledge, which can be broadly classified into two approaches, pre-trained multi-modal embedding and pre-trained user model~\cite{ baltescu2022itemsage,wu-etal-2020-ptum, wu_userbert_2021}. The first approach combines IDRec with pre-trained multi-modal embeddings that are extracted from well-known pre-trained models, such as BERT~\cite{devlin2018bert} and ResNet~\cite{he2016deep}. Subsequently, these modal features are fixed and regarded as supplementary attributes in the recommendation model to alleviate the cold-start problem~\cite{hou2022towards, zhu2021cross}. Despite their success, just using pre-trained representations instead of loading the pre-trained model would greatly limit its ability. Therefore, an alternative approach directly loads the pre-trained user model that is obtained by self-supervision learning from user historical behavior. The final ranking model can simply load the parameters of the pre-trained user model to benefit from the gains brought by large-scale training data.

In spite of the accomplishments yielded by prior investigations on pre-trained knowledge in academic scenarios, the unified deployment between IDRec and the pre-trained user model, still encounters challenges and difficulties in industrial domain. The primary concern is an end-to-end deployment may result in an explosive increase in training resources and online latency, due to the huge amount of the pre-trained model's parameters in industrial recommendation system, an excessive latency may result in vacuous returns, severely impacting the user experience, while users browse the application. Thus, the online latency is  strictly limited within a certain range. Although the method of caching user and item representations in advance is introduced to solve the online latency problem~\cite{liu-etal-2022-boosting}, the performance of the unified model cannot be guaranteed due to lack of the joint training. 

To create a delightful shopping experience, it is critical to accomplish a joint model between IDRec and the pre-trained model without incurring an increase in online latency. Thus, we propose a pre-trained plug-in CTR model (PPM), a novel framework that employs multi-modal features as input and utilizes large-scale data for pre-training. Then, this framework is plugged in IDRec model to mitigate the cold-start problem and accelerate iteration efficiency. Furthermore, during the joint training process, the pre-trained multi-modal features are cached in advance and fixed to address the issue of increased online latency. Meanwhile, the remaining essential parameters of PPM are trainable to further enhance the performance of the unified model. Experimental results show that PPM not only achieves better results on different network structures, but is also effective for cold-start problem. In addition, the performance of utilizing 50\% of data with PPM basically equals that of 100\% without PPM for the unified model, indicating that PPM can significantly improve iteration efficiency for an end-to-end model. 




The contributions of this paper are summarized as follows.
\begin{itemize}
    \setlength\itemsep{0.0em}

    \item We propose a pre-trained plug-in CTR model (PPM) that achieves an end-to-end deployment with IDRec in industrial recommendation system;
    
    

    \item Though caching pre-trained multi-modal features and the other vital parameters of PPM can be optimized during the end-to-end training, the model performance is improved without causing additional online latency increase;
    
    
    \item Experimental results from offline evaluation and online A/B tests prove that PPM not only achieves better results, especially for the cold-start problem, but also can improve iteration efficiency for the unified ranking model.
\end{itemize}

We organize the rest of the paper as follows. In Section~\ref{sec:rw}, we introduce related works in brief. In Section~\ref{sec:method}, we present our proposed method, including PPM and Unified Ranking Model (URM). In Section~\ref{sec:exp} details the experiments, including mentioned questions we care about and relevant experimental results. Finally, we conclude this work in Section~\ref{sec:result}.

\section{Related Work}\label{sec:rw}
\subsection{ID-based Recommendation}
A large number of recommendation models are built only based on discrete User ID and Item ID. Collaborative filtering is a typical method to model user preference by producing similarity matrix based on users' interaction histories, such as Item-based CF~\cite{barkan2016item2vec, sarwar2001item} and User-based CF~\cite{zhao2010user} and Unifing-CF~\cite{wang2006unifying}. Many previous works~\cite{covington2016deep, 48840} simply mapped the sparse IDs into dense vectors, through random initialized embedding matrix. Then deep neural networks are adopted to model users' preference, including DeepFM~\cite{guo2017deepfm}, Wide\&Deep~\cite{cheng2016wide}, GRU4Rec~\cite{hidasi2015session}, SASRec~\cite{kang2018self} and etc~\cite{lee2023mvfs, zhou2019deep}. In the existing recommendation literature, ID-based methods have been well-established and dominated the RS field until now.

\subsection{Pre-trained Modal Embedding}
To solve cold-start problem in IDRec, many prior works adopted two-stage (TS) paradigm to amalgamate IDRec and pre-trained modal embedding. In TS paradigm, the frozen modality representations are extracted from well-known pre-trained models at first. These frozen modality representations are incorporated as auxiliary information for recommendation models, in the form of embeddings~\cite{he2016vbpr, liu2021noninvasive}. ~\citeauthor{he2016vbpr}~\cite{he2016vbpr} attempts to extract visual features using pre-trained deep networks for personalized ranking. ~\citeauthor{zheng2023make}~\cite{zheng2023make} utilize multi-modal encoder and propose a modal adaptation module for better amalgamation of texts and images representations. This approach is popular for industrial applications owing to its cost-effectiveness in terms of computation and training expenses.

\subsection{Pre-trained User Model}
Pre-trained models have made a great success in NLP and CV, which is first pre-trained on unlabeled corpus and then fine-tuned on downstream tasks via labeled data. In recent years, pre-train and fine-tune paradigm is also widely used in recommendation models. 
~\cite{UPRec,wu-etal-2020-ptum} propose pre-training transformer-based recommendation model via predicting the masked item in user behavior sequence. ~\citeauthor{wu_userbert_2021}~\cite{wu_userbert_2021} further proposed contrastive user behavior sequence matching task, combined with masked item prediciton task (MIP) to pre-train user models. 

\section{METHODS}\label{sec:method}
In this section, we describe the architecture of our proposed model as depicted in Figure 1. The whole framework consists of two basic components, Pre-trained CTR Model and Unified Ranking Model. 

\begin{figure*}[htbp]
    \centering
    \includegraphics[width=\linewidth]{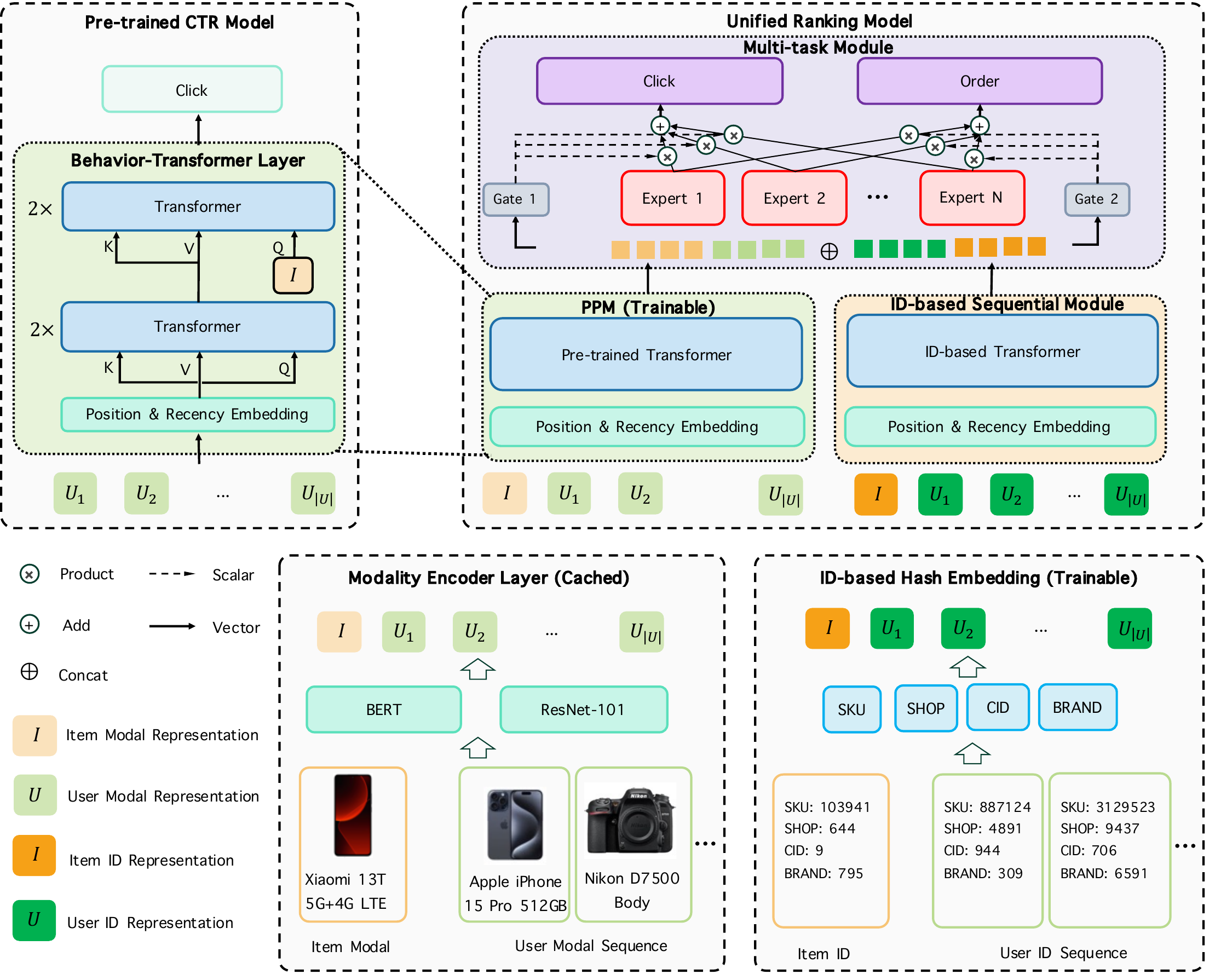}
    \caption{Architecture of the proposed PPM and Unified Ranking Model (URM). }
\label{fig:overall}
\end{figure*}
 
\subsection{Pre-trained CTR Model}
The purpose of pre-trained CTR model mainly focus on building ID-free recommendation model with CTR as the supervision signal, including three layers,  Modality Encoder Layer, Behavior-Transformer Layer and CTR Prediction Layer.

\subsubsection{Modality Encoder Layer} 
The high quality of modality representation(image, text) are obtained by Modality Encoder Layer(ME). In this layer, we fine-tuned pre-trained model(BERT and ResNet) using e-commerce data via query matching task and entity prediction task to obtain the image and text representations for the given items. 
 
\begin{itemize}
\item{\textbf{Query Matching Task}}:
Following SimCSE~\cite{gao2021simcse}, we use contrastive query matching task to train text model with user feedbacks. Let ${x_i, y_i}$ denotes a collection of text-pairs, each pair is formed by a search query and a sequence of textual information of the clicked item from user search catalog. The text-pairs are then encoded by the text model to get embedding-pair $h_{x_i}, h_{y_i}$. The training object is:

\begin{equation}
    \mathcal{L}_{QM} = -\sum_{i}\text{log}\frac{e^{sim(h_{x_i}, h_{y_i})/\tau}}{\sum_{j=1}^{B}e^{sim(h_{x_i}, h_{y_j})/\tau}}
\end{equation}

\noindent where $B$ denotes the batch size, $\tau$ is temperature coefficient.

\item{\textbf{Entity Prediction Task}}:
We use the pre-trained vision model, ResNet-101~\cite{he2016deep} as base model. To adapt general vision model for e-commerce recommendation, we fine-tuned ResNet-101 with product prediction task.  Giving the image of the product $I_i$, our goal is to predict the key entity in the given image $P(E|I_i)$. The training object is:

\begin{equation}
    \mathcal{L}_{EP} = - \sum_{i}\text{log}P(E_i|I_i)
\end{equation}
\end{itemize}

The fixed-length dense features extracted from the frozen text and vision model are concatenated to represent the modality features of the given items.

\subsubsection{Behavior-Transformer Layer} 
To learn the user preference base on the historical interactions between user and items, we first extract the item representation from modality encoder layer, then we utilize bidirectional transformer to capture contextual information, shown in Fig~\ref{fig:overall}(a). Since the transformer layer itself does not contain position information, the transformer in natural language processing uses cosine position encoding. However, the contextual products in the user sequence contain not only sequence information but also time interval information. The time interval between two user interactions may be only 1 second, or even weeks.
 We further use position coding to represent the absolute position of the product displayed on the user interface.
\begin{equation}
E = E_{modal} + E_{position} + E_{recency}
\label{eq:emb}
\end{equation}
\noindent Both $E_{position}$ and $E_{recency}$ are random initialized embeddings. 

Given a sequence of the previous clicked items' modality representation $e_i \in E$ from user catalog $h^u_0$, the contextual representation of item $i$ is calculated through $L^u$ layers of bidirectional transformer blocks:

\begin{equation}
h^u_0=[e_1^u, e_2^u, ..., e_{|\mathcal{S}^u|}^u]
\end{equation}

\begin{equation}
h^u_i=\text{Trm}(h^u_{i-1}, h^u_{i-1}, h^u_{i-1})
\end{equation}

\noindent where Trm denotes bidirectional transformer block, and the input $q$, $k$, $v$ are all $h^u$. $h^u_i$ denotes the output hidden representation at the $i$-th transformer encoder layer. The output of last layer is utilized as user representation in our approach. 

\subsubsection{CTR Prediction Layer} 
By concatenating the representation of user and target item
of the Transformer layer, we then use three fully connected layers to further learn the interactions among the concatenated features, which is standard practice in industrial recomendation systems.

To predict whether a user will click the target item $v_t$, we model it
as a binary classification problem, thus we use the sigmoid function 
as the output unit. To train the model, we use the cross-entropy
loss
\begin{equation}
L=-\frac{1}{N}\sum_{(x,y) \in \mathcal D }{ylogp(x)+(1-y)log(1-p(x))}
\end{equation}

where $\mathcal D$ represent all the samples, and $y \in \{0,1\}$ is the label
representing whether user have click an item or not, $p(x)$ is the
output of the network after the sigmoid unit, representing the
predicted probability of sample $x$ being click.

\subsection{Unified Ranking Model}
The pre-trained CTR model is integrated into IDRec model for end-to-end training. To accelerate the training process and minimize online latency, we initially cache the multi-modal features, enabling the training of the parameters within the Behavior-Transformer Layer and the CTR Prediction Layer, including ID-based sequential module (IDSM), PPM and Multi-task module.


\subsubsection{ID-based Sequential Module}
In the setting of sequential module, we are given an item set $\mathcal I$ and a users interaction sequence $s=\{i_1,i_2,...,i_n\}$ in temporal order where $n$ is the length of $s$ and $i \in \mathcal I$. In IDSM, each item $i$ is represented by side information of ID type, such as stock keeping unit (SKU) id, shop id, brand id and category id, which are all converted to distinct ids and then encoded by item embedding tables. The bi-directional transformer layer (Trm) are utilized to encode user behavior sequences and model the relationship of user and item.  In addition, Trm is not aware of the order of the input sequence. In order to make use of the sequential information of the input, we inject Positional and Recency Embeddings into the input. 

\begin{equation}
E_{ID} = [E_\text{sku\_id}|E_\text{shop\_id}|E_\text{brand\_id}|E_\text{category\_id}] + E_\text{position} + E_\text{recency}
\end{equation}

\noindent $|$ donates concatenation, the output of this module is user interest embeddings and item embeddings based on item's ID features. 

\subsubsection{PPM}
The architecture of the module is the same as ID-based sequential module except the input. In this module, each item $i$ is represented by title and image, which are extracted by ME layer. In order to accelerate the training process and minimize online latency, this representations are cached in hdfs, while other parameters are initialized by preloading the pre-trained CTR model.

\begin{equation}
E_{MO} = [E_\text{title}|E_\text{image}] + E_\text{position} + E_\text{recency}
\end{equation}

The output of this module is  user interest embeddings and item embeddings based on item's multi-modal features.


\subsubsection{Multi-task Module}
Our main goal is to predict the probability of the user clicking, placing an order, or collecting (add to cart) the product. There are both correlations and differences between these tasks. For example, users will definitely click on a product before placing an order, but the products they click on are not necessarily for the purpose of purchase, may just be out of curiosity about a newly released mobile phone. We take advantage of Multi-gate Mixture-of-Experts (MMoE) to model the task relationships and learns task-specific functionalities to leverage shared representations~\cite{ma2018modeling}. Given $N$ experts and $K$ tasks, the output of task $k \in K$ can be formulated as:

\begin{equation}
    f^k(x) = \sum_{i=1}^{n}g^k_i(x)f_i(x)
\end{equation}

\noindent where $f_i(x)$ is the output of $i$-th expert, and $g^k$ denotes the self-gating combination weight of $k$-th task, can be calculated as follow:

\begin{equation}
\begin{aligned}
& g^k = \text{softmax}(xW^k+b^k) \\
& x = [U_{ID}|I_{ID}|U_{MO}|I_{MO}|D]
\end{aligned}
\end{equation}

\noindent where $U_{ID}, I_{ID}$ are the output of IDSM, $U_{MO}, I_{MO}$ are the output of PPM and $D$ is other features. $x \in \mathbb{R}^N$, $W^k \in \mathbb{R}^{N \times N}$ and $b^k \in \mathbb{R}^N$ are learnable weight matrix and bias vector of the gating module.

After calculating all gating networks and experts, we can obtain the prediction of task k finally:

\begin{equation}
    y^k(x) = t^k(f^k(x))
\end{equation}

\noindent where $t_k$ denotes the tower network of task k.

\section{Experiments}\label{sec:exp}

In this section, we first describe experimental settings and then conduct extensive experiments to answer the following research questions:
\begin{itemize}
\item{\textbf{RQ1:}} How do the various state-of-art sequential recommendation models enhanced with the proposed PPM perform?
\item{\textbf{RQ2:}} How does each component of PPM perform in predicting the click and order tasks? 
\item{\textbf{RQ3:}} Does the proposed URM advantage over the traditional CTR model without PPM in mitigating the cold-start issue?
\end{itemize}

\subsection{Dataset}

We leverage real-world production CTR datasets. Each sample in the datasets contains a sequence of user historical behaviors, the target item, and the ground truth label indicating whether or not the user clicked the target item. We use these datasets to evaluate the effectiveness of our proposed model. We first used about 6 months of data for model pre-training. Then, 1 weeks (Small) and 2 weeks (Large) of data were used for training the Unified Ranking model respectively. The details of the three training datasets are shown in Table~\ref{tab:dataset}.

\begin{table}[htb]
    \centering
    \begin{tabular}{cccc} \hline 
         Dataset & \#Interactions &  \#User &  \#Items  \\ \hline 
         Pretrain & 12,418,618,242 & 505,439,104 & 93,322,811 \\ 
         Small & 546,746,823 & 113,482,552 & 23,888,962  \\
         Large & 1,295,355,022 & 163,660,999 & 38,502,046 \\ \hline 
    \end{tabular}
    \caption{Statistics of the CTR datasets used in the experiments.}
    \label{tab:dataset}
\end{table}

\subsection{Experimental Settings}
\subsubsection{Evaluation metrics}
Following the previous work~\cite{shen2022deep,zheng2022hien,zhu2021open}, we evaluate the ranking result using two widely used metrics: AUC and Precision@N.
\begin{itemize}
\item{\textbf{AUC}(Area Under the ROC Curve)}: AUC represents the probability that a random positive example is positioned in front of a random negative example. The higher, the better.
\item{\textbf{Precision@N}}: In an information retrieval system that retrieves a ranked list, the top-N documents are the first N in the ranking. Precision at N is the proportion of the top-N documents that are relevant~\cite{craswell_precision_2009}.
\end{itemize}

In the online A/B test, we also employ UCTR (clicks per user per day) and UCVR (orders per user per day) to evaluate the performance of our model.

\begin{equation}
\begin{aligned}
& UCTR = \frac{\#clicks}{\#users} \\
& UCVR = \frac{\#orders}{\#users}
\end{aligned}
\end{equation}

\subsubsection{Compared Methods}
We compare the proposed approach with the following baseline methods in Table ~\ref{tab:allperformance}. In this paper, we focus on comparing our method with existing classical user interest modeling methods that mainly model dynamic user behavior from historical interactions.
\begin{itemize}
    \item{\textbf{Wide\&Deep}~\cite{cheng2016wide}}: Wide\&Deep (W\&D) trains a wide linear model and a deep neural model simultaneously for CTR prediction, combining the benefits of memorization and generalization.
    \item{\textbf{DeepFM}~\cite{guo2017deepfm}}: To combine the power of traditional FM and deep MLP, DeepFM replaces the LR in the wide network of Wide \&Deep with FM to model the 2-order feature interactions.
    \item{\textbf{DIN}~\cite{richardson2007predicting}}: DIN is a deep model that employs an attentive neural network to activate related user behaviors with respect to corresponding targets.
    \item{\textbf{DSIN}~\cite{feng2019deep}}: This is a transformer-based model that uses the transformer and RNN to model the user’s intra- and intersession interests separately.
    \item{\textbf{MIAN}}~\cite{zhang2021multi}: MIAN is a deep CTR model that contains a multi-interaction and transformer layer to extract multiple representations of user behavior.
\end{itemize}

\begin{table*}[htb]
	\centering
	\scalebox{0.95}{\begin{tabular}{c|l|cccccc|cccccc}
			\toprule 

        \multirow{3}{*}{\textbf{Dataset}} &  \multirow{3}{*}{\textbf{Model}}&
   \multicolumn{6}{c}{\textbf{wo.PPM}}&	\multicolumn{6}{|c}{\textbf{w.PPM}} \\
   &&	\multicolumn{2}{|c}{\textbf{Click}}& \multicolumn{2}{c}{\textbf{Order}}& \multicolumn{2}{c}{\textbf{Average}} &	\multicolumn{2}{|c}{\textbf{Click}}& \multicolumn{2}{c}{\textbf{Order}} &\multicolumn{2}{c}{\textbf{Average}} \\
    
   & & AUC & P@2 & AUC & P@2 & $\overline{AUC}$ & $\overline{P@2}$ & AUC & P@2 & AUC & P@2  & $\overline{AUC}$(Improv) & $\overline{P@2}$(Improv) \\
   \midrule 
            \multirow{5}{*}{Small}  
& Wide\&Deep & 0.6094 & 0.1886 & 0.6030 & 0.1655 & 0.6062 & 0.1770 & 0.6797 & 0.2029 & 0.6948 & 0.1685 & \textbf{0.6873}(0.0811) & \textbf{0.1857}(0.0087)\\
&DeepFM & 0.6051 & 0.1882 & 0.6138 & 0.1688 & 0.6095 & 0.1785 & 0.6784 & 0.2028 & 0.6920 & 0.1686 & \textbf{0.6852}(0.0758) & \textbf{0.1857}(0.0072)\\
&DIN & 0.6946 & 0.2168 & 0.7231 & 0.2013 & 0.7089 & 0.2091 & 0.7093 & 0.2242 & 0.7451 & 0.2099 & \textbf{0.7272}(0.0183) & \textbf{0.2171}(0.0080)\\
&DSIN & 0.7000 & 0.2189 & 0.7332 & 0.2055 & 0.7166 & 0.2122 & 0.7107 & 0.2241 & 0.7422 & 0.2084 & \textbf{0.7265}(0.0098) & \textbf{0.2163}(0.0041)\\
&MAIN & 0.6859 & 0.2170 & 0.7174 & 0.1973 & 0.7016 & 0.2072 & 0.6926 & 0.2225 & 0.7279 & 0.2030 & \textbf{0.7102}(0.0086) & \textbf{0.2127}(0.0056)\\
&URM & 0.7228 & 0.2283 & 0.7648 & 0.2301 & 0.7438 & 0.2292 & 0.7279 & 0.2324 & 0.7676 & 0.2325 & \textbf{0.7477}(0.0040) & \textbf{0.2324}(0.0023)\\

\midrule 
\multirow{5}{*}{Large} 
&Wide\&Deep & 0.6011 & 0.1883 & 0.6129 & 0.1668 & 0.6070 & 0.1775 & 0.6848 & 0.2065 & 0.7021 & 0.1734 & \textbf{0.6935}(0.0865) & \textbf{0.1899}(0.0124)\\
&DeepFM & 0.6046 & 0.1875 & 0.6233 & 0.1749 & 0.6140 & 0.1812 & 0.6853 & 0.2069 & 0.7024 & 0.1737 & \textbf{0.6939}(0.0799) & \textbf{0.1903}(0.0091)\\
&DIN & 0.7038 & 0.2208 & 0.7373 & 0.2063 & 0.7205 & 0.2135 & 0.7130 & 0.2261 & 0.7482 & 0.2106 & \textbf{0.7306}(0.0100) & \textbf{0.2184}(0.0048)\\
&DSIN & 0.7069 & 0.2232 & 0.7416 & 0.2087 & 0.7243 & 0.2159 & 0.7125 & 0.2254 & 0.7464 & 0.2107 & \textbf{0.7295}(0.0052) & \textbf{0.2181}(0.0021)\\
&MAIN & 0.6945 & 0.2215 & 0.7266 & 0.2014 & 0.7106 & 0.2114 & 0.6996 & 0.2251 & 0.7287 & 0.2038 & \textbf{0.7141}(0.0036) & \textbf{0.2145}(0.0030)\\
&URM & 0.7279 & 0.2326 & 0.7685 & 0.2323 & 0.7482 & 0.2324 & 0.7343 & 0.2363 & 0.7722 & 0.2313 & \textbf{0.7532}(0.0050) & \textbf{0.2338}(0.0014)\\

			\bottomrule 
	\end{tabular}}
	\caption{The overall performance comparison with other baseline methods}
    \label{tab:allperformance}
\end{table*}

\subsubsection{Implementation Details}
All models are trained on GPUs(Tesla A100) using Tensorflow2. The batch size is set to 20000. The learning rates for Adam~\cite{kingma2014adam} with $\eta$ = 0.0004 to 0.0001. The model parameters are initialized with a Gaussian
distribution (with a mean of 0 and a standard deviation of 0.01). The id embedding and multi-modal embedding dimensions are set to 64. The layer number of all transformer-encoder and transformer-decoder is 2 and the number of multi-attention heads is 4. 

\subsection{Performance Comparison (RQ1)}
Table~\ref{tab:allperformance}  reports the experimental results of the baseline methods and our proposed model on real-world small and large CTR datasets. The best results are highlighted in boldface. Note that the improvement is the absolute AUC/P@2 improvement of models with PPM to the models without PPM. It can be observed that all recommendation models enhanced with PPM outperform the models without PPM, which demonstrates the effectiveness of our proposed model. Besides, we also have the following observations:

\begin{itemize}
    \item Among the conventional methods, W\&D and DeepFM perform the worst. In contrast, attention-based methods have better performance for CTR prediction. This verifies the necessity of modeling contextual representation of user behavior sequence. 
    
    \item DSIN outperforms DIN and MAIN, owing to its effectiveness in extracting users’ historical behaviors into session interests and modeling the dynamic evolution of session interests.
    
    \item PPM model exhibits a greater improvement on the small dataset compared to the large dataset, further indicating that the enhancement brought by PPM model is more pronounced in scenarios with sparse data.
    


    \item Except for URM, the effects of other models trained with PPM on a small dataset outperform the models without PPM trained on a large dataset. The effects are the essentially identical between small dataset and large dataset on URM (In our scenario, the deviations within 0.1\% are fluctuations). This indicates that PPM can greatly accelerate the iteration efficiency of URM.
    
\end{itemize}


\begin{table*}
    \centering
    \scalebox{0.95}{\begin{tabular}{l|cc|cc|cc}
			\toprule 

         \multirow{2}{*}{\textbf{Model}}&  \multicolumn{2}{|c}{\textbf{Click}}& \multicolumn{2}{|c}{\textbf{Order}}& \multicolumn{2}{|c}{\textbf{Average}} \\
   & AUC & P@2  & AUC & P@2  & $\overline{AUC}$ (Improv) & $\overline{P@2}$ (Improv)  \\
   \midrule 
      Base & 0.7252 & 0.2310 & 0.7612 & 0.2305 & 0.7432 (-) & 0.2308 (-)\\
Base+QM\&EP & 0.7279 & 0.2326 & 0.7685 & 0.2323 & 0.7482 (0.0050) & 0.2324 (0.0017)\\
Base+QM\&EP+PPM (random initialized) & 0.7277 & 0.2321 & \textbf{0.7735} & \textbf{0.2351} & 0.7506 (0.0074) & 0.2336 (0.0028)\\
Base+QM\&EP+PPM (frozen) & 0.7318 & 0.2347 & 0.7710 & 0.2317 & 0.7514 (0.0082) & 0.2332 (0.0024)\\
Base+QM\&EP+PPM (finetune) & \textbf{0.7343} & \textbf{0.2363} & 0.7722 & 0.2313 & \textbf{0.7532} (0.0100) & \textbf{0.2338} (0.0031)\\
    \bottomrule 
	\end{tabular}}
    \caption{The performance of contrast models in terms of AUC and P@2 for click and order tasks.}
    \label{tab:ablation}
\end{table*}

\subsection{Ablation Study (RQ2)}
In this section, we conduct experiments to evaluate the effectiveness of different components in URM. Specifically, we design four contrast models:
\begin{itemize}
    \item{Base}: URM without Query Matching Task and Entity Prediction Task (QM\&EP), Pre-trained Plug-in Model (PPM). Thus, the multi-modal features are extracted from primal pre-trained BERT and ResNet by self-supervised learning. Then, the multi-modal features are concatenated with other ID-based features and encoded through a shared transformer.
    \item{Base+QM\&EP}: Base model with QM\&EP (URM wo.PPM).
    \item{Base+QM\&EP+PPM (random initialized)}: URM with random initialized Plug-in Model.
    \item{Base+QM\&EP+PPM (frozen)}: URM with frozen Pre-trained Plug-in Model.
    \item{Base+QM\&EP+PPM (finetune)}: This is our proposed unified ranking model (URM).
\end{itemize}

Table~\ref{tab:ablation} shows the performance of contrast models in terms of AUC and P@2 for click and order tasks. URM achieves the best performance in terms of AUC and P@2, which proves the effectiveness of each component of our model. Furthermore, we also have the following observations:

\begin{itemize}
    \item The model with QM\&EP significantly outperforms the base model, which proves that fine-tuning modality encoder models with QM\&EP tasks play an indispensable part in PPM. Fine-tuned BERT and ResNet offer more appropriate representations for text and images in the e-commerce domain.
    \item We notice that the model with random initialized  PPM outperform others without PPM, indicating that modeling ID-based features and modal-based features with separate transformers is an optimal choice in our scenario.
    \item We discover that models with either frozen or fine-tuned PPM outperform the model with random initialized PPM, which suggests the effectiveness of our proposed pre-trained method for PPM. 
    \item We also observe that frozen PPM can degrade the performance of PPM, this may be because PPM needs to adapt for domain-speciﬁc data and learn user representations specialized for downstream tasks.
\end{itemize}

\begin{figure}[htbp]
    \centering
    \includegraphics[width=\linewidth]{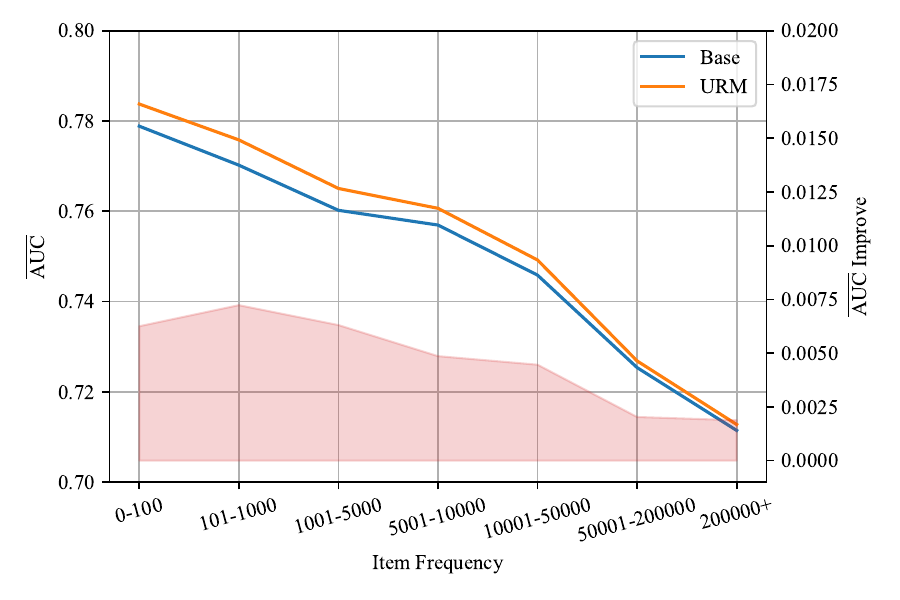}
    \caption{The performances of our proposed URM and Base model in cold-start and warm settings. The shaded region in red represents the observed increase in $\overline{AUC}$ for URM compared to the Base Model.}
    \label{fig:cold}
\end{figure}

\subsection{Performance in Cold Start Setting (RQ3)}
Fig~\ref{fig:cold} illustrates the performances of our proposed URM and Base model in different settings. Both models are trained on a large dataset. We partitioned the test dataset into different groups based on the appearance frequency of the target item in the training dataset. The lower the appearance frequency of the target item, the more long-tail the test sample represents. The results demonstrate that our URM surpasses the Base model, particularly in long-tail items, thereby validating the effectiveness of URM in mitigating the cold-start issue through the proposed PPM.

\subsection{Online A/B Testing}
The A/B test is conducted on the JD homepage recommendation service for 10 consecutive days, where the baseline model is our last online CTR model which the modality features are regarded as a kind of side information for item. Simultaneously, to make the online evaluation confident and fair, each method deployed for the A/B test has the same number of users. PPM contributes to a 1.09\% increase in UCTR, a 0.28\% increase in UCVR, and promotes a 10\% increase in the exposure ratio of long-tail products. Now PPM has been deployed online and serves the main traffic of users. 

Figure 3 demonstrates the workflow of the deployed URM, including offline training and incremental updates. During offline training, we utilize large-scale data to train PPM. Then, 50\% of the data, which are utilized to train traditional rank model without PPM, are employed to train URM with PPM. In order to ensure the timeliness of the ranking model, we timely update both PPM and URM. PPM is updated by using the past 2 days (T-2) and URM is updated by using the past 1 day (T-1) with the last PPM. After incremental updates, URM is deployed online to ensure users' shopping experience.


\begin{figure}[htbp]
    \centering
    \includegraphics[width=\linewidth]{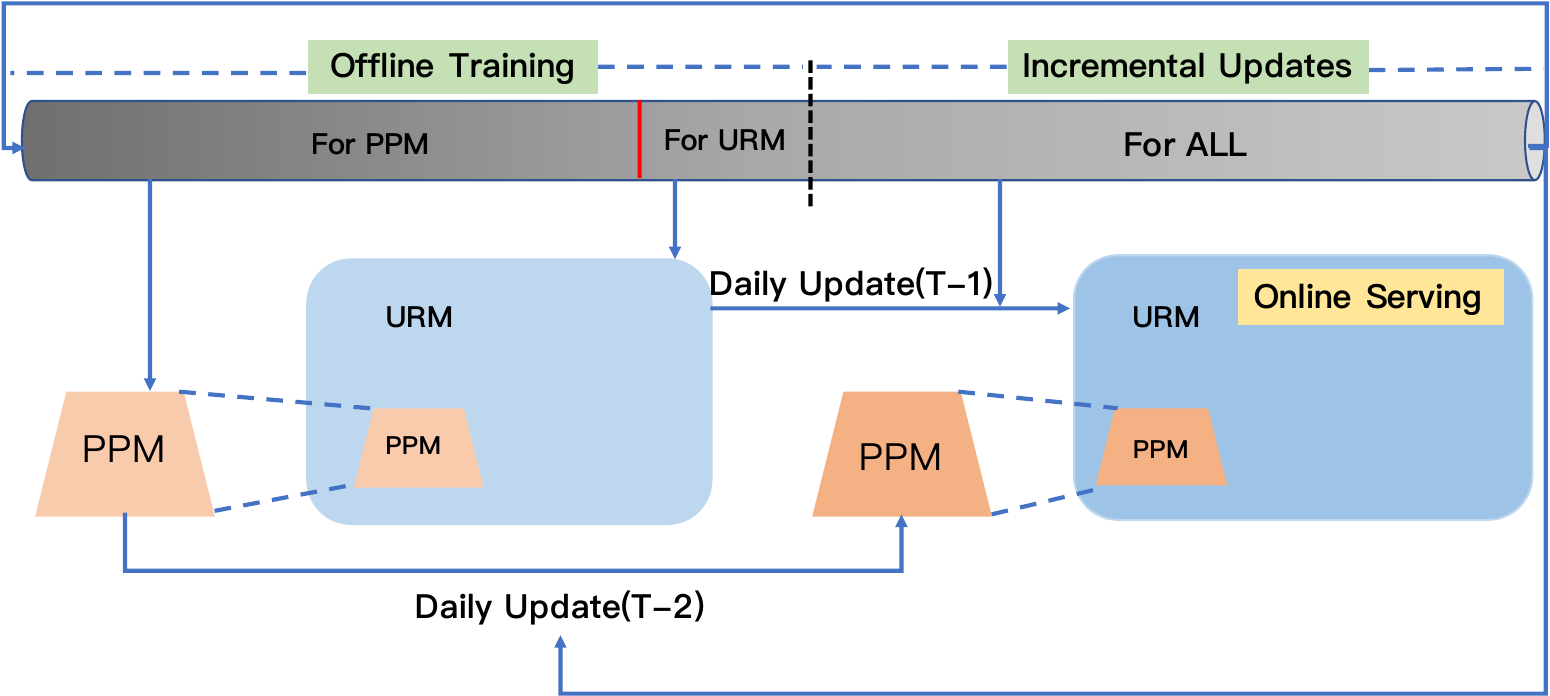}
    \caption{The offline training and incremental update processes of PPM and URM.}
    \label{fig:pipeline}
\end{figure}

\section{Conclusion}\label{sec:result}
In this paper, we propose a pre-trained plug-in CTR model (PPM) to boost the performance of IDRec in industrial recommender systems. PPM is built on the well-known pre-trained model, utilizing multi-modal features (title and image of item) as input and CTR as a supervision signal for pre-training at first. Subsequently, it is loaded by the IDRec model and only a subset of crucial parameters to be trained concurrently. Both offline and online A/B testing results demonstrate the effectiveness of our approach without an increase in online latency.


\bibliographystyle{ACM-Reference-Format}
\balance
\bibliography{sample-base}

\end{document}